\magnification=1200
\parskip 10pt
\parindent 12pt
\baselineskip=18pt
\input mssymb
\pageno=0
\footline={\ifnum \pageno <1 \else \hss \folio \hss \fi }
\line{\hfil{DAMTP-R/94/1}}
\line{\hfil{January, 1994}}
\vskip 1in
\centerline{{\bf THE FOUR-POINT FUNCTION ON A SURFACE OF INFINITE GENUS}
{\footnote{*}{This article has appeared in Modern Physics Letters A
${\underline 9}$(14) (1994)}}}
\vskip .8in
\centerline{Simon Davis}
\vskip .5in
\centerline{Department of Applied Mathematics and Theoretical Physics}
\vskip 1pt
\centerline{University of Cambridge}
\vskip 1pt
\centerline{Silver Street, Cambridge CB3 9EW}
\vskip .6in
{\bf Abstract}. The four-point function arising in the scattering of closed
bosonic strings in their tachyonic ground state is evaluated on a surface of
infinite genus. The amplitude has poles corresponding to physical intermediate
states and divergences at the boundary of moduli space, but no new types of
divergences result from the infinite number of handles.  The implications for
the universal moduli space approach to string theory are briefly discussed.
\vfill
\eject

The geometry of string perturbation theory suggests that a more complete
formulation might be achieved by using the analyticity and connectedness
properties of a universal moduli space [1] or Grassmannian [2] to treat the
amplitudes at arbitrary genus uniformly.  Both spaces contain points corresponding to
infinite-genus surfaces, and it would be useful to establish if calculations
of correlation functions and scattering amplitudes can be extended to this
order.

The integration of the positions of vertex operators over world sheets
corresponding to surfaces of infinite genus could involve unphysical
divergences as a result of the non-compactness of the manifold.  The four-point 
amplitude for closed bosonic strings is computed here when the surface is a sphere
with an infinite number of handles which become infinitesimally small (Fig. 1).

\input epsf.tex

\vbox{
\epsfysize=2.4in
\centerline{\epsfbox{sphere.eps}}
\vskip 0.2in
\noindent {\bf Fig. 1.  A sphere with an infinite number of handles.  The
distance
between the handles and their size decreases to zero.}}

It will be shown that the amplitude is finite, except for poles in the momenta,
which occur at values corresponding to physical intermediate states, and
divergences associated with the boundary of moduli space.  The result depends
on the rate at which spacing between the handles and their size decreases to
zero.

Since the scattering amplitudes for closed bosonic strings depends on the
correlation functions of vertex operators, the Green function for the scalar
Laplacian on a Riemann surface is required when the incoming and outgoing
strings are in their tachyonic ground states.  Although the Green function
on a compact surface may be expressed in terms of prime forms, it also can be
obtained by using the representation, up to conformal equivalence of the
surface
as $D/\Gamma$, where D is a domain in the extended complex plane and $\Gamma$
is
a discontinuous subgroup of $PSL(2,{\Bbb C})$ leaving D invariant [3], and then
applying
the method of images.  In particular, uniformization of a closed surface of
genus g by a Schottky group generated by the linear transformations
$T_1,...,T_g$ leads to an expression for the Green function with two sources
located at the points $z_R$ and $z_S$
$$\eqalign{G_{QS}(P,R)~=~\sum_\alpha&~ln~\left\vert {{z_P~-~V_\alpha z_R}\over
{z_P~-~V_\alpha z_S}} {{z_Q~-~V_\alpha z_S}\over {z_Q~-~V_\alpha
z_R}}\right\vert
\cr
-~& {1\over {2 \pi}}~\sum_{m,n=1}^g~Re \{v_m (z_P)~-~v_m
(z_Q)\}~(Im~\tau)_{mn}^{-1}~Re\{v_n (z_R)~-~v_n (z_S)\}
\cr
v_n(z)~&=~\sum_\alpha~^{(n)}~ln\left({{z~-~V_\alpha \xi_{1n}}\over
{z~-~V_\alpha
\xi_{2n}}}\right)~~~~~~v_n (z)~-~v_n(T_m z)~=~ 2 \pi i~\tau_{mn}
\cr}
\eqno(1)$$
where the $V_\alpha$ are arbitrary products of the generators $T_1,...,T_g$,
$\xi_{1n}, \xi_{2n}$ are the two fixed points of $T_n$, $\sum_\alpha^{(n)}$
represents the sum over all $V_\alpha$ that do not have $T_n^{\pm 1}$ at the
right-hand end of the product, and $\tau$ is the period matrix of the
surface.  The finiteness of the sums in equation (1) depends on the convergence
of the Poincare series $\sum_{\alpha \not= I}~\vert \gamma_\alpha \vert^{-2}$
[4], where
$V_\alpha z~=~ {{\alpha_\alpha z+\beta_\alpha}\over {\gamma_\alpha z +
\delta_\alpha}}$.

For the extension of the Schottky group to an infinite number of generators,
the isometric circles $I_{T_n}~=~\{z \in \hat{\Bbb C} \vert \vert\gamma_n z
+ \delta_n\vert ~=~1\}$ can be joined to $I_{T_n^{- 1}}~=~\{z \in \hat{\Bbb C}
\vert \vert \gamma_n z - \alpha_n\vert~=~1\}$ for all n to create a sphere with
an infinite number of handles [5].  While the expression for the Green function
in terms of prime forms is no longer available, because the theta function has
been generalized only for particular infinite-genus surfaces [6], the method of
images can be used again to obtain a Green function in the form (1).

A new proof of convergence of the Poincare series is necessary when the
uniformizing group $\Gamma$ is infinitely generated.  Since $\sum_{\alpha \not=I}
\vert \gamma_\alpha \vert^{-2}~=~ \sum_{\alpha \not= I} ~r_\alpha^2$, where
$r_\alpha$ is the radius of $I_{V_\alpha}$, the sum can be shown to be finite
when $r_n$ decreases to zero sufficiently fast as $n \to \infty$ [7].  For
instance, suppose that the distances between the isometric circles are bounded below,
and that the distance between $I_{T_n}$ and $I_{T_n^{-1}}$ is bounded above for
all n, so that the circles accumulate at $\infty$ as in Fig. 2.  Then, if
$r_n~=~ kn^{-2}$, with k being a constant that depends on the spacing between the 
circles, the series converges.  This result can be demonstrated by grouping the
elements of $\Gamma$ according to the number of fundamental generators in the product.
Let $V_{(l)}$ be an element of the Schottky group at level $l$ consisting of
the product of $l$ fundamental generators.  Suppose that $V_{(l+1)}~=~T_n~V_{(l)}$.
Then
$$\left\vert{{\gamma_{(l+1)}}\over {\gamma_{(l)}}}\right\vert~>~
(\vert K_n\vert^{-{1\over 2}}~-~\vert K_n\vert^{1\over 2}) \left\vert
{{\xi_{2n}~-~{{\alpha_{(l)}}\over {\gamma_{(l)}}}}\over {\xi_{2n}~-~\xi_{1n}}}
\right\vert~-~\vert K_n\vert^{1\over 2}
\eqno(2)$$
If the decrease of the absolute values of the multipliers is given by $\vert
K_n\vert^{1\over 2}~=~(c_1 n^2+c_2)^{-1}$, the ratio in equation (2) is greater
than $c_1c n^2$ when $c_2^2~>~1+{1\over c}$, where c is the lower bound for
$\left\vert {{\xi_{2n}~-~{{\alpha_{(l)}}\over {\gamma_{(l)}}} }\over
{\xi_{2n}~-~
\xi_{(1n)}}} \right\vert$.  Denoting the upper bound for $\vert
\xi_{2n}~-~\xi_{1n}\vert$ by $c^\prime$, it follows that
$\vert \gamma_n\vert^{-2}~<~{{c^{\prime 2}}\over {c_1^2 n^4}}$ and
$$\sum_{\alpha \not= I}~\vert \gamma_\alpha \vert^{-2}~<~ c^2c^{\prime 2}
\left[{2\over {c_1^2 c^2}}~\sum_{n=1}^{\infty}~{1\over {n^4}}~+~
\left({2\over {c_1^2 c^2}}~\sum_{n=1}^{\infty}~{1\over {n^4}}\right)^2~+~...
~\right]
\eqno(3)$$
which converges when $c_1^2~>~ {2\over {c^2}}~\sum_{n=1}^{\infty}~{1\over
{n^4}}$.  In terms of the radii of the isometric circles of the fundamental
generators, the decrease is given by $k n^{-2}$, where $k~<~{{c^{\prime}}\over
{c_1}}$.

While convergence of the Poincare series implies finiteness of the first sum in
equation (1), the second term becomes
\hfil\break
$-{1\over {2 \pi}} \sum_{m,n=1}^{\infty}
{}~Re\{v_m (z_P)~-~v_m (z_Q)\} (Im ~\tau)_{mn}^{-1} Re\{v_n (z_R)~-~v_n
(z_S)\}$
at $g=\infty$.  This sum is well-defined when the
imaginary part of the period matrix is positive-definite, a property which
follows from the bilinear relations for harmonic differentials on a
Riemann surface of class $O_G$ [8][9].  In particular, the inverse of the
period matrix exists for the class of surfaces considered above.
{}From the formula relating the entries of the period
matrix and the Schottky group parameters
$$\tau_{mn}~=~{1\over {2 \pi i}}~\left[~ln~K_m~\delta_{mn}~+~
               \sum_\alpha~^{(m,n)}~ln~\left({{\xi_{1m}~-~V_\alpha \xi_{1n}}
               \over {\xi_{1m}~-~V_\alpha \xi_{2n}}}
              {{\xi_{2m}~-~V_\alpha \xi_{2n}}\over
                   {\xi_{2m}~-~V_\alpha \xi_{1n}}}\right)~\right]
\eqno(4)$$
where the elements $V_\alpha$ having $T_m^{\pm 1}$ as the left-most member and
$T_n^{\pm 1}$ as right-most member of the product are excluded from the
summation, it follows that
$$Im~\tau_{nn}~\simeq~{2\over \pi}~ln~n~+~{1\over \pi}~ln~c_1~+~ {{c_2}\over
                              {\pi c_1 n^2}}
\eqno(5)$$
Both the off-diagonal elements of $Im~\tau$ and the functions $Re~v_n(z)$ can
be estimated simultaneously.  Consider a point z at a bounded distance $d(z,
I_{T_{n_0}})$
from $I_{T_{n_0}}$ for finite $n_0$.  Since
$$Re~v_n (z)~=~\sum_\alpha~^{(n)} ~ln~\left\vert~1~+~{{(\xi_{2n}~-~\xi_{1n})
\gamma_\alpha^{-2}}\over {(\xi_{1n}~+~{{\delta_\alpha}\over {\gamma_\alpha}})
(\xi_{2n}~+~{{\delta_\alpha}\over {\gamma_\alpha}})(z~-~V_\alpha \xi_{2n})}}
\right\vert~\equiv~\sum_\alpha~^{(n)}~Re~ v_{n \alpha}(z)
\eqno(6)$$
the elements $V_\alpha$ can be separated into the following categories:
$$\eqalign{(i)~~~& d(I_{V_\alpha}, I_{T_{n_0}}),~ d(I_{V_\alpha^{-1}},
      I_{T_{n_0}})~ are~ bounded
\cr
&Re~v_{n \alpha} (z)~=~O\left({1\over {n^2}}\right)~\vert
\gamma_\alpha\vert^{-2}
\cr
(ii)~~~& d(I_{V_\alpha}, I_{T_{n_0}}),~d(I_{V_\alpha^{-1}},
I_{T_n})~are~bounded
\cr
&Re~v_{n \alpha}(z)~=~O \left({1\over {n^3}}\right)
{}~\vert \gamma_\alpha\vert^{-2}
\cr
(iii)~~~& d(I_{V_\alpha}, I_{T_n}),~d(I_{V_\alpha^{-1}},
I_{T_{n_0}})~are~bounded
\cr
&Re~v_{n \alpha} (z)~=~O(1) ~\vert \gamma_\alpha \vert^{-2}
\cr
(iv)~~~& d(I_{V_\alpha}, I_{T_n}),~d(I_{V_\alpha^{-1}}, I_{T_n})~are~bounded
\cr
&Re~v_{n \alpha} (z)~=~O \left({1\over n} \right)~\vert \gamma_\alpha
\vert^{-2}
\cr}
\eqno(7)$$
The sum over the elements in the categories (iii) and (iv) may be estimated by
repeating the analysis of the Poincare series $\sum_{\alpha \not= I} \vert
\gamma_\alpha \vert^{-2}$ with the restriction $I_{V_\alpha} \subset
D_{T_m},~D_{T_m^{-1}} $
for $m \geq n$, where $D_{T_m}$ is the isometric disk $D_{T_m}~=~\{z \in
\hat{\Bbb C}\vert \vert \gamma_m z+\delta_m \leq 1\}$.
The sum can be bounded by
$$2{{c^{\prime 2}}\over {c_1^2}}~{{\zeta(4,n)}\over {1~-~\zeta(4,n)}}
\eqno(8)$$
with $\zeta (4,n)$ being the generalized zeta function.  It may be verified
that a lower bound with similar dependence on n can be placed on the
restricted Poincare series.  Using Hermite's representation of the generalized
zeta function, one observes that the bound (8) decreases as ${{2 c^{\prime 2}}
\over {3 c_1^2}}{1\over {n^3}}$.
Since the contributions of the elements in categories (i) and (ii) decrease
as $O({1\over {n^2}})$ and $O({1\over {n^3}})$ respectively, $Re~v_n (z)~=~
O({1\over {n^2}})$. Thus, $Re~\{v_n (z_P)~-~v_n (z_Q)\}~ < ~{{v_{PQ}}\over
{n^2}}$ for an appropriate constant $v_{PQ}$.
The fall-offs of the entries $(Im ~\tau)_{mn},~m \not= n$ are then
$$Im \tau_{mn}~=~O \left({1\over {\vert m ~-~ n\vert^2}}\right)
\eqno(9)$$
Since diagonalization of the matrix $(Im \tau)^{-1}$ produces eigenvalues
$\lambda_n~=~{\pi \over 2}{1\over {ln~ n}}$ for large n,
\noindent
${1\over {2 \pi}}~\sum_{m,n=1}^{\infty}~Re~\{v_m (z_P)~-~v_m (z_Q)\}
(Im~\tau)^{-1}_{mn}~ Re~\{v_n (z_R)~-~v_n (z_S)\}
\hfil\break
{}~<~{1\over 4}
{}~v_{PQ}~v_{RS}~ \sum_{n=1}^{\infty}~{1\over {n^4~ln~n}}$
so that this term is also finite confirming the results of a general proof
given in earlier work [7].

Having established the suitability of the series expansion (1) for the Green
function, one would like to determine its behaviour near the isometric circles,
particularly in the region where they accumulate at $\infty$, because the
scattering amplitude for N tachyons
$$\eqalign{\sum_{g=0}^{\infty}~&\kappa^g ~\int_{M_g}~ d \mu_g
{}~\int_{\Sigma_g}~\prod_{i=1}^N~d^2
z_i~\sqrt{g(z_i)}~\prod_{i<j}~e^{-{{p_i\cdot
p_j}\over 2}~\tilde G^{sym}(z_i, z_j)}
\cr
\tilde G^{sym} &(z_i, z_j)~=~G^{sym}(z_i, z_j)~-~{1\over 2} lim_{z_i^{\prime}
\to z_i}~[G(z_i, z_i^{\prime})~-~ln~ d(z_i, z_i^{\prime})]
\cr
.&~~~~~~~~~~~~~~~~~~~~~~~~-~{1\over 2} lim_{z_j^{\prime} \to z_j}~[G(z_j,
z_j^{\prime})~-~ln~d(z_j, z_j^{\prime})]
\cr}
\eqno(10)$$
involves an integration over the entire fundamental domain, $\Delta$ of the
Schottky group,  whose points are in one-to-one correspondence with the
Riemann surface.

Consider, for example, the four-point amplitude with the positions of three of
the vertex operators fixed at $z_1^0$, $z_2^0$, $z_3^0$
$$\eqalign{f(z_1^0, z_2^0, z_3^0)~\int_\Delta&~d^2 z_4~ \vert
z_4~-~z_1^0\vert^{-{{p_1 \cdot p_4}\over 2}}~\prod_{\alpha \not=
I}~\left\vert{{z_4~-~V_\alpha z_1^0}\over
{z_4~-~V_\alpha z_4}}{{z_1^0~-~V_\alpha z_4}\over {z_1^0~-~V_\alpha z_1^0}}
\right\vert^{-{{p_1 \cdot p_4}\over 4}}
\cr
\cdot\prod_{m,n}&~exp~\left[-{{p_1 \cdot p_4}\over {8 \pi}} Re(v_m (z_4)~-~v_m
(z_1^0)) (Im~\tau)^{-1}_{mn} Re(v_n(z_4)~-~v_n(z_1^0))\right]
\cr
\cdot &(similar~factors~ with~z_1^0~\to~z_2^0, z_3^0,~p_1~\to~p_2,p_3)
\cr}
\eqno(11)$$
where
$$\eqalign{f(z_1^0, z_2^0, z_3^0)~=&~\vert z_2^0~-~z_1^0\vert^{2-{{p_1 \cdot
p_2}\over 2}}
\prod_{\alpha \not= I}~\left\vert{{z_2^0~-~V_\alpha z_1^0}\over
{z_2^0~-~V_\alpha z_2^0}}{{z_1^0~-~V_\alpha z_2^0}\over {z_1^0~-~V_\alpha
z_1^0}}\right\vert^{-{{p_1 \cdot p_2}\over 4}}
\cr
\times \prod_{m,n}&~exp [(-{{p_1 \cdot p_2}\over {8 \pi}}
{}~Re(v_m (z_2^0)~-~v_m (z_1^0))~ (Im \tau)^{-1}_{mn}~Re(v_n
(z_2^0)~-~v_n(z_1^0))]
\cr
\times & (similar~ factors ~involving ~permutations~ of~z_1^0, ~z_2^0, ~z_3^0)
\cr}
\eqno(12)$$
The integral (11) is finite in the neighbourhood of $z_1^0,~z_2^0,~z_3^0$
if $p_1 \cdot p_4,~p_2 \cdot p_4,~p_3 \cdot p_4 < 4$, while
momentum conservation implies that $p_1 \cdot p_4 ~+~p_2 \cdot p_4,~p_1 \cdot
p_4~+~p_3 \cdot p_4,~p_2 \cdot p_4~+~p_3~+~\cdot p_4~>4$.  This demonstrates
the existence of a range of momenta for which the integral is finite, allowing
for analytic continuation of the amplitude to physical values of the momenta.

The integral in the neighbourhood of the isometric circles can be studied to
ascertain whether the allowed range is narrowed even further.  Specifically,
the integral is finite in the asymptotic regions $z_4~\to~-{{\delta_\alpha}
\over {\gamma_\alpha}},~z_4~\to~V_\alpha z_1^0,~ z_4~\to~V_\alpha z_4$, and
$z_4~\to~\infty$, by the property of momentum conservation.  The product
terms in (11) conceivably could diverge if $V_\alpha z_1^0$ or $V_\alpha z_4$
is arbitrarily close to one of the circles $I_{T_n}$.  This does not happen,
however, because, there are no limit points on any of the isometric circles
$\{I_{T_n}\}$ [7].  Moreover, for the configuration in Fig. 2, if z is a
bounded distance away from $I_{T_n}$, the maximum distance between 
$V_\alpha z$ and the center of $I_{T_n^{-1}}$ is ${{c_0 r_n}\over n}$ 
for some constant $c_0$.

This result can be obtained by considering the positions of the isometric
circles $I_{(T_n T_m)}^{-1}$ inside the disk $D_{T_n^{-1}}$.  Suppose that
z lies on the curve defined by the equation
$${{\vert \gamma_m\vert}\over {\vert \gamma_n \vert}}{{\vert
z~-~{{\alpha_m}\over{\gamma_m}}\vert }\over {\vert z~+~{{\delta_n}\over
{\gamma_n}}\vert}}~=~1
\eqno(13)$$
Then $T_n z$ lies on $I_{(T_n T_m)^{-1}}$ and its distance from the center of
$D_{T_n^{-1}}$ is
$$\left\vert T_n z~-~{{\alpha_n}\over {\gamma_n}} \right\vert~=~
{{\vert \gamma_n \vert^{-2}}\over {\left\vert z~+~ {{\delta_n}\over {\gamma_n}}
\right\vert}}~=~{{\vert \gamma_n\vert^{-1} \vert \gamma_m \vert^{-1}}
\over {\left \vert z~-~ {{\alpha_m}\over {\gamma_m}}\right\vert}}
\eqno(14)$$
For $m~\ll n$,
$$d(I_{(T_n T_m)^{-1}}, I_{T_n^{-1}})~\simeq \vert\gamma_n \vert^{-1}
\left\vert 1~-~{1\over {m^2 n}} \right\vert
\eqno(15)$$
It follows that
$$\vert z_4~-~V_\alpha z_1^0 \vert~>~ \vert \gamma_n \vert^{-1}
\left(1~-~{{c_0}
\over n}\right)
\eqno(16)$$
for some value of $c_0$.  Similarly, $\vert z_4~-~V_\alpha z_4 \vert~<~
\vert \gamma_n \vert^{-1} \left(1~+~{{c_0}\over n} \right)$ when $z_4$ lies in
an infinitesimal neighbourhood of $I_{T_n^{-1}}$.
\vskip 0.2in
\input epsf.tex

\vbox{
\epsfysize=2.7in
\centerline{\epsfbox{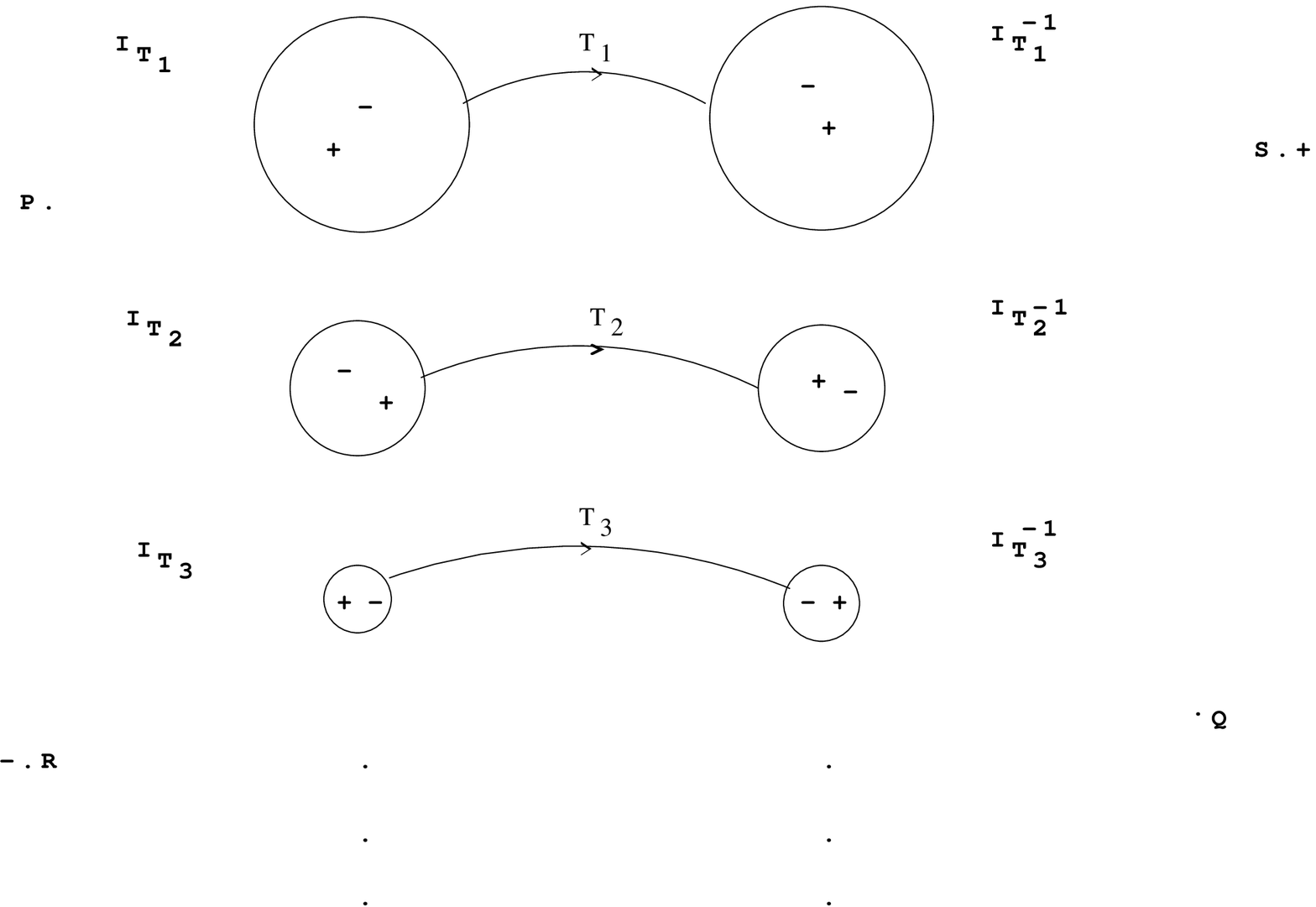}}
\vskip 0.2in
\noindent{\bf Fig. 2.  The fundamental region for an infinitely discontinuous
subgroup
of PSL(2, ${\Bbb C}$) is the exterior of the isometric circles.}}

One can then place bounds such as ${{1-{{c_0}\over n}}\over {1+{{c_0}\over n}}}
{}~<~\left\vert {{z_4 -V_\alpha z_1^0}\over {z_4-V_\alpha z_4}}\right\vert~<~
{{1+{{c_0}\over n}}\over {1-{{c_0}\over n}}}$, which imply that the product
factors in (11) tend to 1 as $z_4$ approaches $\infty$ along the direction of
the isometric circles.  Since $-2{{c_0}\over n}-O({1\over
{n^2}})~\leq~ln~\left\vert{{z-V_\alpha \xi_{1m}}\over {z-V_\alpha
\xi_{2m}}}\right\vert~\leq~2{{c_0}\over n} +O({1\over {n^2}})$ when $V_\alpha
z$
lies in $D_{T_n}$ or $D_{T_n^{-1}}$, the contribution of the isometric circles
accumulating at $\infty$ to $Re~v_n (z)$ as $z \to \infty$ vanishes.   As the
contribution of the elements $V_\alpha$, such that $I_{V_\alpha^{-1}} \subset
D_{T_{n_0}}, D_{T_{n_0}^{-1}}$ for bounded $n_0$ also vanishes as $z \to
\infty$, $Re~v_n (z) \to 0$ in this limit.
 Therefore, since $\sum_i p_i~=~0,~p_4^2~=~-8$, the
integrand falls off as $\vert z_4 \vert^{-4}$, and the integral is finite.
For any value of $s,~t,~u$, the only singularities in the integrand occur at
$z_1^0,~z_2^0,~z_3^0$ so that (11) is of the form
$$f(z_1^0,z_2^0,z_3^0)~\int_\Delta d^2 z_4~\vert z_4~-~z_1^0 \vert^{-{{p_1
\cdot
p_4}\over 2}}~\vert z_4~-~z_2^0\vert^{-{{p_2 \cdot p_4}\over 2}}
{}~\vert z_4~-~z_3^0 \vert^{-{{p_3 \cdot p_4}\over 2}}~ \Phi(z_4, {\bar {z_4}})
\eqno(17)$$
where $\Phi(z_4, {\bar {z_4}})$ is regular throughout the fundamental region.
By dividing the integration region into three disks of radius $\Lambda$ about
$z_1^0,~z_2^0,~z_3^0$ and the remainder of the fundamental domain, the
integrand can be expanded in a Laurent series [7][10] and the integral can be
shown to be equal to
$$\eqalign{2 \pi &\sum_{n=0}^{\infty} ~{{\Lambda^{-{{p_1 \cdot p_4}\over
2}+2n+2}}
\over {-{1\over 2} p_1 \cdot p_4+2n+2}}{1\over  {(n!)^2}}
(\partial {\bar\partial})^n \{\vert z_4~ -~z_2^0 \vert^{-{{p_2 \cdot p_4}\over
2}}
\vert z_4~-~z_3^0 \vert^{-{{p_3 \cdot p_4}\over 2}} \Phi(z_4, {\bar
{z_4}})\}_{z_4=z_1^0}
\cr
+&(similar ~terms~ with~z_1^0 \to z_2^0, z_3^0, p_1 \to p_2, p_3)~+~finite
\cr}
\eqno(18)$$
In terms of the Mandelstam variables, there are simple poles at s, t, u =
8(n-1), n = 0, 1, 2, ..., corresponding to the tachyon and excited
intermediate states.

Although the positions of three of the vertex operators are fixed in (11), the
full amplitude requires an integration over $z_1^0,~z_2^0,~z_3^0$ also.  The
analysis above still holds, except that are singularities in the product terms
when $z_i$ approaches $I_{T_n}$ and $I_{T_n^{-1}}$ for any n.  These
divergences
are physical, however, because $z_i$ and $z_j$ are approaching the same point
on the world sheet.

The total scattering amplitude, given by a further integration over moduli
space,  would also contain divergences associated with the boundary of moduli
space.  For the class of surfaces defined by $\vert \xi_{1n}~-~ \xi_{2n} \vert$
being bounded and $\vert K_n \vert~=~(c_1 n^2~+~c_2)^{-2}$, there is a
divergence in the moduli space integral given by
$${{sinh^4 \left(\sqrt{{c_2}\over {c_1}} \pi \right)}\over {\pi^4 c_2^2}}
lim_{g \to \infty} {{c_1^{4g+2} \pi^{14 g - 4}}\over {2^{13 g}}}
(g!)^8 \prod_{n=1}^g~ {1\over {(ln~n)^{13}}}
\eqno(19)$$
up to exponential factors associated with products over conjugacy classes of
primitive elements and fixed-point integrals [11][12].

This investigation has demonstrated that these are the only types of
divergences
that arise in the amplitude even though the surfaces have an infinite number of
handles.  This result could be extended to the scattering of other string
states
such as the graviton.  The derivatives in the corresponding vertex operator
would be expected to lead to an even faster fall-off for the correlation
functions as the positions tend to infinity along the direction of the
isometric
circles.  Finiteness of the integral of the correlation function over the
string worldsheet, except for coincidence of the vertex operators, would follow
from this asymptotic behaviour.

The growth of the moduli space integral given in equation (19) is similar to
the the dependence of the partition function, regularized in a
genus-independent
manner by introducing cut-offs on the lengths of closed geodesics on the
Riemann surface [13].  However, the divergence in (19) represents an independent
contribution to the moduli space integral, which would be eliminated by the
regularization, because there exist closed geodesics, homotopically equivalent
to the $A_n$-cycles, or $I_{T_n}$ on the covering surface, that have length in
the intrinsic metric
decreasing to zero as $n \to \infty$ [7][14].

Divergences at the boundary of moduli space can be eliminated in superstring
theory at each finite order of the perturbation expansion [15].  Since the
amplitudes at each order have been demonstrated to be finite, no regularization
removing a neighbourhood of the boundary of moduli space is required.
Effectively closed surfaces of the type considered in this paper could be
included in the path integral representing the scattering amplitude.
The large-order divergences found for bosonic strings may be eliminated for
superstrings, since it has been shown that they arise in the Schottky group
parametrization in the limits $\vert K_n \vert \to 0$ and $\vert \xi_{1n}~-~
\xi_{2n}\vert \to 0$ [14].  However, as a larger class of surfaces is being
included
in the superstring path integral, the counting of the different types of
surfaces would affect finiteness of the entire scattering amplitude.
\vskip 10pt

\centerline{\bf Acknowledgements.}
\noindent I would like to thank Prof. C. Vafa for useful discussions on string
theory.  I have benefitted from conversations with Dr. M. Awada,  Prof. D. Kazhdan,
Prof. I. Kra, Prof. B. Maskit, Dr. C. Nunez and Prof. A. Verjovsky.  
Part of this work initially appeared as a
Harvard preprint HUTP-88/A002.  I am grateful to Prof. S. Coleman for his encouragement
of the beginning of this research at the Lyman Laboratory, where it was
supported by an S.E.R.C. NATO Fellowship and in part by NSF grant PHY-82-15249.
The interest of Dr. G. W. Gibbons and Prof. S. W. Hawking in this research
is also gratefully acknowledged.

\vfill
\eject
\centerline{REFERENCES}

\item{[1]}  D. Friedan and S. Shenker, Phys. Lett. ${\underline{B175}}$ (1986)
287 \hfil\break
D. Friedan and S. Shenker, Nucl. Phys. ${\underline{B281}}$ (1987) 509

\item{[2]}  N. Ishibashi, Y. Matsuo and H. Ooguri, University of Tokyo preprint
UT-499(1986)
\hfil\break
L. Alvarez-Gaume, C. Gomez and C. Reina, Phys. Lett. ${\underline{B190}}$
(1987) 55
\hfil\break
 E. Witten, Commun. Math. Phys. ${\underline{113}}$ (1988) 529 - 600

\item{[3]}  J. Lehner,
${\underline{Discontinuous~Groups~and~Automorphic~Functions}}$,
\hfil\break
 Mathematical
Surveys No. 8 (Providence: American Mathematical Society, 1964)

\item{[4]}  W. Burnside, Proc. Lond. Math. Soc. Vol. 23 (1892) 49

\item{[5]}  B. Maskit, ${\underline{Kleinian~Groups}}$ (Berlin:
Springer-Verlag, 1990)

\item{[6]}  H. P. McKean and E. Trubowitz, Commun. Pure Appl. Math. Vol. XXIX
(1976) 143

\item{[7]}  S. Davis, Class. Quantum Grav. ${\underline{6}}$ (1989) 1791 - 1803

\item{[8]}  L. Sario and M. Nakai,
${\underline{Classification~Theory~of~Riemann~
Surfaces}}$
(Berlin:
\hfil\break
Springer-Verlag, 1970)

\item{[9]}  R. D. M. Accola, Trans. Amer. Math. Soc. ${\underline{96}}$ (1960)
143
\hfil\break
Y.Kusonoki, Mem. Coll. Sci. Univ. Kyoto Series A Math. ${\underline{30}}$(1)
(1956) 1 - 30

\item{[10]}  G. Aldazabal, M. Bonini, R. Iengo and C. Nunez, ICTP  preprint
 IC/87/319

\item{[11]}  S. Davis, Class. Quantum Grav. ${\underline{7}}$ (1990) 1887 -
1893

\item{[12]}  A. Erdelyi, W. Magnus, F. Oberhettinger, F. G. Tricomi,
\hfil\break
${\underline{Higher~Transcendental~Functions:~Bateman~Manuscript~Project}}$,
Vol. 1
\hfil\break
(New York: McGraw
Hill, 1953)

\item{[13]}  D. J. Gross and V. Periwal, Phys. Rev. Lett. ${\underline{60}}$
(1988) 2105 - 2108

\item{[14]}  S. Davis, ICTP preprint IC/92/431

\item{[15]}  N. Berkovits, Nucl. Phys. ${\underline{B408}}$ (1993) 43-61

\end